\documentstyle[12pt,epsf]{article}
\newcommand {\beq}{\begin{equation}}
\newcommand {\eeq}{\end{equation}}
\newcommand {\beqa}{\begin{eqnarray}}
\newcommand {\eeqa}{\end{eqnarray}}
\newcommand {\n}{\nonumber \\}

\begin{document}
\setlength{\oddsidemargin}{0cm}
\setlength{\baselineskip}{7mm}

\begin{titlepage}
 \renewcommand{\thefootnote}{\fnsymbol{footnote}}
$\mbox{ }$
%\vspace{-3cm}
\begin{flushright}
\begin{tabular}{l}
IMSC-2000-08- \\
hep-th/0008019\\
August 2000
\end{tabular}
\end{flushright}

~~\\
~~\\
~~\\
\vspace*{0cm}
    \begin{Large}
       \vspace{2cm}
       \begin{center}
         {Electric-Magnetic Duality in IIB Matrix Model with D-brane$^{}$\footnote
{I am sorry there may be a lot of strange English in this preprint. 
I will correct them as soon as possible
and show as revised version.
}
}
\\
       \end{center}
    \end{Large}

  \vspace{1cm}

\begin{center}

                        Hiroyuki T{\sc akata}$^{}$\footnote
           {
e-mail address : takata@imsc.ernet.in}

        $^{}$ {\it The Institute of Mathematical Sciences (IMSc),}\\
               {\it C.I.T Campus, Chennai 600 113, Taramani, India}\\
\end{center}
%Latex, 15 pages, 1 figure, 
%talk at Lattice 2000, 17-22 August 2000, Bangalore, India

\vfill
%%%%%%%%%%%%%%%%%%%%%%%%%%%%%%%%%%%%%%%%%%%%%%%%%%%%%%%%%%%%%%%%%%%%%%%%%%%%%%%%%%%%%%%%
\begin{abstract}
\noindent
We consider electric-magnetic duality(S-duality) 
in IIB matrix model with a D3-brane background. 
We propose the duality transformation by considering that of 
noncommutative Yang-Mills theory(NCYM) in four dimension. 
NCYM derived from the matrix model has a Yang-Mills coupling 
related to the noncommutativity of the spacetime. 
We argue that 
open strings bits as bi-local fields on the spacetime 
are decoupled from the bulk in NCOS decoupling limits 
as it is in string theory approach. 
We also discuss 
how our four dimensional spacetime relates to higher, 
by applying the decoupling and the commutative limits 
of the backgrounds of the matrix model.
\end{abstract}
%%%%%%%%%%%%%%%%%%%%%%%%%%%%%%%%%%%%%%%%%%%%%%%%%%%%%%%%%%%%%%%%%%%%%%%%%%%%%%%%%%%%%%%%

\vfill
\end{titlepage}
\vfil\eject
%%%%%%%%%%%%%%%%%%%%%%%%%%%%%%%%%%%%%%%%%%%%%%%%%%%%%%%%%%%%%%%%%%%%%%%%%%%%%%%%%%%%%%%%
%%%%%%%%%%%%%%%%%%%%%%%%%%%%%%%%%%%%%%%%%%%%%%%%%%%%%%%%%%%%%%%%%%%%%%%%%%%%%%%%%%%%%%%%
%%%%%%%%%%%%%%%%%%%%%%%%%%%%%%%%%%%%%%%%%%%%%%%%%%%%%%%%%%%%%%%%%%%%%%%%%%%%%%%%%%%%%%%%
\section{Introduction}
\setcounter{equation}{0}

Our world is a brane$^{}$\footnote
           {``brane'' here is some space 
which has an almost definite the spacetime dimension, 
and in particular, 
means fundamental string or Dirichlet brane in string theory.}; 
this picture gives us a direct understanding 
of nature from string theory.
This idea is so attractive 
and a lot of people are trying to derive 
our spacetime 
from some consistent theory - string theory$^{}$\footnote
{``string theory'' in this paper means the theory of string described by world sheet.
``string'' is, however, used with the matrix model as well as ``string theory''.
}
 or its matrix description
\cite{maldacena}\cite{AIKKT}\cite{GKP}\cite{Witten}\cite{RS}\cite{NV}.
Unfortunately, we have not succeeded 
to get neither the spacetime dimension 4,
gauge group, the variety of quark masses etc.
Deriving the number 4 may be the easiest.
Although these tasks are  not done in this paper,
the motivation is along the line.

%%%%%%%%%%%%%%%%%%%%%%%%%%%%%%%%%%%%%%%%%%%%%%%%%%%%%%%%%%%%%%%%%%%%%%%%%%%%%%%%%%%%%%%%
The electric-magnetic duality is originally a duality 
of exchanging electronic (${\cal E}$) and magnetic (${\cal B}$) field
in Maxwell equations in 4 dimension$^{}$\footnote
{Euclid version are considered in this paper} \cite{Maxwell}\cite{Dirac}:
\beqa
\mbox{\bf div}{\cal E}= \rho_e \,\,,&& \mbox{\bf rot}{\cal B}-\dot{{\cal E}}=j_e\,\,,\n
\mbox{\bf div}{\cal B}= \rho_b \,\,,&& \mbox{\bf rot}{\cal E}-\dot{{\cal B}}=j_b\,\,.
\label{Maxwell}
\eeqa
It has been extended to that of supersymmetric YM 
and superstring theory as S-duality \cite{MO}\cite{Sen}\cite{SW1994}.
4 dimensional spacetime as D3-brane naturally appears 
in type IIB superstring theory\cite{Polchinski}.
Since this whole system is the selfdual, 
the duality symmetry gives us some information.  
Now we should think of the duality 
on noncommutative 4 dimensional spacetime.
We will concentrate, however, 
only to electric-magnetic duality of noncommutative U(1).
In the case of noncommutative U(1),
the dual transformation exchanges 
the spacespace non commutativity $\theta^{01}$ 
and the spacetime noncommutativity $\theta^{23}$.
This is because the action includes terms 
contracted over Lorentzian indices among
${\cal E}$, ${\cal B}$ and $\theta^{01}$, $\theta^{23}$\cite{RU}\cite{GRS}.

%%%%%%%%%%%%%%%%%%%%%%%%%%%%%%%%%%%%%%%%%%%%%%%%%%%%%%%%%%%%%%%%%%%%%%%%%%%%%%%%%%%%%%%%
Noncommutative Yang-Mills theory(NCYM) appears 
in string theory with D-brane.
The coordinates parameterizing D-brane world volume  are, 
in generic, non commutative\cite{Bachas}\cite{Cheung:1998nr}.
Then it is natural to replace the coordinates by matrix valued ones.
The traditional string theory have only 1-branes within a 9-brane,
then the dynamical variables are written like this:
\beq
X(\tau, \sigma)\,\,,
\eeq
where both $X$ and $\tau, \sigma$ are just $1 \times 1$ matrices.
Now there are various dimensions of brane in string theory, 
and lower dimensional branes live there in general. 
So we may write roughly,
\beq
\cdots \left(X_p
      \left(X_{p^\prime}
      \left(X_{p^{\prime \prime}}
      \left( \cdots
      \right)
      \right)
      \right)
      \right)\cdots \,\, , \,\,\,
         \left(\cdots p \geq p^\prime \geq p^{\prime \prime} \cdots \right)\,\,,
\eeq
where $X_p$ means a collection of coordinates of branes 
having lower dimension than $p+1$, 
and all $X_p$'s  are matrix valued.   
It is too complicated. 
While there is some consistency condition 
simplifying it\cite{K},
IIB matrix model\cite{IKKT}\cite{review} 
is assumed to describe this brane complex.
The dynamical variables are 10 matrices(and their superpartners),
whose each entry is , surprisingly, just a number 
independent of any parameters.

This simplification was first found 
as Eguchi-Kawai reduction\cite{EK}\cite{Parisi} 
in the gauge theory framework,
and found for branes\cite{Townsend}\cite{Ishibashi}.  
In order to compensate of the reduction 
of the degrees of freedom, 
the size of those matrices n 
is assumed to be large enough$^{}$\footnote
{In this paper we do not answer how large n and do not care about finite n corrections}.
These equivalences have been understood 
as Morita equivalence recently\cite{CDS}\cite{PS}.
In fact NCYM4 can be derived from the matrix model 
as an example.
This is further studied in\cite{bars99}\cite{AMNS}\cite{AMNS1}.
Our future plan is to understand 
where informations of NCYM and string theory 
are embedded in  the matrix model. 

%%%%%%%%%%%%%%%%%%%%%%%%%%%%%%%%%%%%%%%%%%%%%%%%%%%%%%%%%%%%%%%%%%%%%%%%%%%%%%%%%%%%%%%%
In IIB matrix model, 
diagonal components of matrices 
describe the relative coordinates of spacetime points 
and off-diagonal ones correspond interactions among them.
So if these coordinates have non zero value 
only for 4 direction,
it means our 4 dimensional universe. 
Actually, we can find 3-brane classical solution of the action.
This 3-brane has a NCYM4 on it\cite{AIIKKT}, 
where gauge fields are from the quantum fluctuation 
while spacetime is from the classical background.  
Since the 3-brane solution forms Heisenberg algebra,
4 dimensional space is constructed as  Hilbert space.
Although spacetime coordinate has uncertainty relation,
the coherent states make a intuitive 4 dimensional spacetime
as von Neumann lattice\cite{IKK}\cite{IIKK}\cite{GN}.
In the paper\cite{IKK}\cite{IIKK}, 
they showed open strings on the lattice 
and relates the noncommutativity to the string scale.  

%%%%%%%%%%%%%%%%%%%%%%%%%%%%%%%%%%%%%%%%%%%%%%%%%%%%%%%%%%%%%%%%%%%%%%%%%%%%%%%%%%%%%%%%
In approach to NCYM 
from the open string 
in the presence of the NS B field\cite{SW98},
there is a decoupling limit 
of open strings on the brane from the bulk( NCOS limit)\cite{SST}\cite{GMMS}.
The electric field cannot be larger than a critical value.
It is because of the Lorentzian metric.
In the  critical value, 
open strings in the world volume become tensionless 
and are decoupled from closed strings in  bulk.    
This feature suggests the existence 
of a non-critical string theory only with open strings.
Rather, it means a possibility of constructing 
some lower dimensional spacetime from 10 dimensional one. 

In the matrix model framework, 
one can understand this simply.
In the paper\cite{KT}, 
we have found the correspondence 
between the  electromagnetic field ($\tilde{E}$, $\tilde{B}$) 
on the open strings
and the spacetime noncommutativity($\theta^{01}$, $\theta^{23}$) 
in the matrix model.
Namely$^{}$\footnote
{In the paper\cite{KT}, 
we compared in Lorentzian.
},
\beqa
(1-\tilde{E}^2){\theta^{01}}^2& = &({2 \pi \alpha^\prime})^2\,\,,\n
(1+\tilde{B}^2){\theta^{23}}^2& = &({2 \pi \alpha^\prime)}^2\,\,.
\label{KT}
\eeqa
This relation tells us 
what is the decouple limits in the matrix model as we will see.
We will get the intuitive explanation 
by using open strings bits on the von Neumann lattice. 
It is also consistent to the feature in string theory\cite{SST}\cite{GMMS}.,
This is directly understood when we rewrite the eq.(\ref{KT}) to eq.(\ref{thetas}),
which is a relation 
between the noncommutativity $\theta^{01}$ in the matrix model 
and $\theta^{01}_{s}$ in string theory derived in Ref.\cite{SW98}\cite{SST}.

%%%%%%%%%%%%%%%%%%%
In this paper, we propose a electromagnetic duality 
of a 3-brane spacetime-time constructed from the matrix model.
And the NCOS decoupling limit is found there.

%%%%%%%%%%%%%%%%%%%%%%%%
In section 2,   
NCU(1)4 from the matrix model are derived. 
Where, non-self dual a 3-brane solution are treated,
namely, $\theta^{01} \not\equiv \theta^{23}$.
Since we  are interested in the relation to the case of commutative 
and the case of decoupled spacetime,
we treat non selfdual solution 
of background matrices in the matrix model.
We also see open strings in it in terms of bi-local fields.
Each direction of momenta of open strings 
depends on the corresponding noncommutativity. 
The gauge coupling $g_{YM}$ 
and noncommutativity $\theta^{01}$, $\theta^{23}$ of the spacetime
are related\cite{AIIKKT} as eq.(\ref{gYM}):
\beq
g^2_{YM} \sim  \theta^{01}\theta^{23}.
\eeq     

%%%%%%%%%%%%%%%%%%%%%%%%%%
In section 3, 
the electric-magnetic duality are considered.
The standard prescription of dual transformation 
is formulated as Legendre transformation 
in partition function\cite{HL}.
We use the relation of\cite{RU} for our NCU(1) case.
Then the dual representation of 
the field  strength, 
the YM coupling 
and the noncommutativity of NCU(1)4 are gotten; eq.(\ref{dual}) and eq.(\ref{dualg}).
The U(1) coupling $g_{YM}$ cannot to be 1 by rescaling of the gauge field ${\cal A}$ in noncommutative case. 
In the matrix model framework, however, it is possible by rescaling of matrices.
Thus we get the duality of Maxwell equation in a 4 dimensional commutative spacetime. 

%%%%%%%%%%%%%%%%%%%%%%%%%%%%%%%%%%%%%%%%%%%%%%%%%%%%%%%%%%%%%%%%%%%%%%%%%%%%%%%%%%%%%%%%
In section 4,
NCOS decoupling limit is defined 
in the matrix model( large $\theta^{\mu \nu}$ limit).
The correspondence of different approaches, 
that is, from open string with the B field 
and from IIB matrix model
are gotten in Ref.\cite{KT} and eq(\ref{KT}) .   
The momenta of open strings bits on the von Neumann lattice
are small near the limit; see eq.(\ref{momenta}).
This means the open strings cannot form loops 
and decouple from closed strings.
Rather, 4 dimensional spacetime are decoupled 
from other transverse directions.
On the other hand, 
commutative limits( small $\theta^{\mu \nu}$ limit) can also be taken.
It is opposite limit to the decoupling limit.
Different limits are chosen for different directions,
since $\theta^{\mu \nu} \not\equiv \theta^{\rho \sigma}$ now.
Thus, we will propose how our 4 dimensional spacetime happens from higher. 

%%%%%%%%%%%%%%%%%%%%%%%%%%%%%%%%%%%%%%%%%%%%%%%%%%%%%%%%%%%%%%%%%%%%%%%%%%%%%%%%%%%%%%%%
Section 5
includes discussions of dynamical generation of 4 dimensional spacetime.

%%%%%%%%%%%%%%%%%%%%%%%%%%%%%%%%%%%%%%%%%%%%%%%%%%%%%%%%%%%%%%%%%%%%%%%%%%%%%%%%%%%%%%%%%
%%%%%%%%%%%%%%%%%%%%%%%%%%%%%%%%%%%%%%%%%%%%%%%%%%%%%%%%%%%%%%%%%%%%%%%%%%%%%%%%%%%%%%%%%
%%%%%%%%%%%%%%%%%%%%%%%%%%%%%%%%%%%%%%%%%%%%%%%%%%%%%%%%%%%%%%%%%%%%%%%%%%%%%%%%%%%%%%%%%

\section{Noncommutative U(1) theory in four dimension}
\setcounter{equation}{0}

In this section
NCU(1)4 is derived from the matrix model.
In order to see the duality of the matrix model,
we first look into a theory around it, 
that is, NCU(1).

We start by following IIB matrix model action\cite{IKKT}\cite{AIIKKT}:
\beq
S=- {1\over g^2}Tr({1\over4}[A_{\mu},A_{\nu}][A_{\mu},A_{\nu}]
+{1\over 2}\bar{\psi}\Gamma _{\mu}[A_{\mu},\psi ]) .
\label{IIBaction}
\eeq
Now $A_\mu$ and $\psi$ are $n\times n$ Hermitian matrices
and each component of $\psi$ is 10 dimensional Majorana-spinor.
We expand $A_\alpha=\hat{p}_\alpha+ \hat{\cal A}_\alpha$, 
(for $\alpha=0,1,2,3$)around the following classical solution
\beq
\left[\hat{p}_\alpha,\hat{p}_\beta\right]=iF_{\alpha \beta}
\eeq
\beq
F_{\alpha \beta}=\left( \begin{array}{cccc}
0 & -1/\theta^{01} &0 &0 \\
1/\theta^{01} &0 &0 &0\\
0 &0 &0 &-1/\theta^{23} \\
0&0 & 1/\theta^{23}&0 
\end{array}
\right) ,
\label{sol}
\eeq
where $\theta^{01}$, $\theta^{23}$ are c-numbers.
$\theta $ is defined as inverse of $F$.
The noncommutative coordinates are introduced as:
\beq
\hat{x}^\alpha:=\theta^{\alpha \beta}\hat{p}_\beta\,\,,
\eeq
and satisfy the relation:
\beq
[\hat{x}^\alpha,\hat{x}^\beta]=-i \theta^{\alpha \beta}\,\,.
\label{xx}
\eeq
Since we are going to see the duality and the decoupling limit,
$\theta$ may not be self dual, namely, $\theta^{01} \not\equiv \theta^{23}$.

Followed Ref.\cite{AIIKKT}\cite{IKK}\cite{IIKK},
$\hat{\phi}:=\{ \hat{\cal A}_\alpha, \hat{\varphi}_i:=A_i, \hat{\psi} \}$, 
($\alpha,\beta =0 \sim 3, i,j = 4 \sim 9 $)
are mapped to usual functions on phase space formed by noncommutative coordinates explicitly:
\beq
\hat{\phi} \rightarrow \phi(x)=\sum_k \tilde{\phi}_k e^{i k_\alpha x^\alpha}\,\,.
\eeq
The summation over $k_\alpha$ is performed as follows\cite{AIIKKT}:
\beqa
k_{\alpha=0,1}=
\pm{1 \over n^{1 \over 4}}\sqrt{2\pi \over |\theta^{01}|}\,\,,
\pm{2 \over n^{1 \over 4}}\sqrt{2\pi \over |\theta^{01}|}\,\,,
\cdots\,\,,
\pm{n^{1 \over 4} \over n^{1 \over 4}}\sqrt{2\pi \over |\theta^{01}|}\,\,,&&\n
\pm{n^{1 \over 4}+1 \over n^{1 \over 4}}\sqrt{2\pi \over |\theta^{01}|}\,\,,
\cdots\,\,,
\pm{n^{1 \over 4} \over 2}\sqrt{2\pi \over |\theta^{01}|}\,\,,&&\n
k_{\alpha=2,3}=
\pm{1 \over n^{1 \over 4}}\sqrt{2\pi \over |\theta^{23}|}\,\,,
\pm{2 \over n^{1 \over 4}}\sqrt{2\pi \over |\theta^{23}|}\,\,,
\cdots\,\,,
\pm{n^{1 \over 4} \over n^{1 \over 4}}\sqrt{2\pi \over |\theta^{23}|}\,\,,&&\n
\pm{n^{1 \over 4}+1 \over n^{1 \over 4}}\sqrt{2\pi \over |\theta^{23}|}\,\,,
\cdots\,\,,
\pm{n^{1 \over 4} \over 2}\sqrt{2\pi \over |\theta^{23}|}\,\,.&&
\label{momenta}
\eeqa
Then we get the action of NCU(1) from eq.(\ref{IIBaction}):
\beqa
S_{\bf NCU(1)}&=&{1\over (2\pi g)^2 \theta^{01}\theta^{23}}
\int d^{4}x
\Big({1\over 4}{\cal F}_{\alpha \beta}{\cal F}_{\alpha \beta}\n
&&+{1\over 2}[D_{\alpha},\varphi_i][D_{\alpha},\varphi_i]
+{1\over 4}[\varphi_i,\varphi_j][\varphi_i,\varphi_j]\n
&& -{i \over 2} \bar{\psi}{\Gamma}_{\alpha}[D_{\alpha},\psi ]
- {1 \over 2} \bar{\psi}\Gamma_i[\varphi_i,\psi ]
\Big)_{\star} .
\eeqa
Inside $(\mbox{\hspace{3mm}})_\star$, the products should be understood as:
\beqa
(\phi_1 \phi_2)_\star(x)&:=& \phi_1(x)\star\phi_2(x)\n
&:=&exp({\theta^{\alpha \beta} \over 2 i}
{\partial^2 \over \partial \xi^{\alpha}\partial \eta^{\beta}} )
(\phi_1(x + \xi) \phi_2(x+\eta)|_{\xi=\eta=0}
\label{star}
\eeqa 
The covariant derivative  and the field strength are defined as:
\beq
D_\alpha:=\partial_\alpha -i {\cal  A}_\alpha\,\,,\,\,\,\, 
{\cal F}_{\alpha\beta}:=i[D_{\alpha},D_{\beta}]_\star
\eeq
The Yang-Mills coupling is related to the noncommutativity as:
\beq
g^2_{YM}=(2\pi g)^2 \theta^{01}\theta^{23}\,\,.
\label{gYM}
\eeq
Now eq.(\ref{Maxwell}) is the non commutative Maxwell equations,
where ($l,m,n=1\sim3$) 
\beqa
\rho_e &:=& [\varphi_i, \dot{\varphi}_i]_\star 
+ \{ \bar{\psi}_a, \psi_b \}_\star (\Gamma_0)_{ab}\,\,,
\rho_b=0\,\,,\n
j_e &:=& [\varphi_i , {\bf grad \varphi_i}]_\star 
+ \{ \bar{\psi}_a, \psi_b \}_\star ( {\bf \Gamma})_{ab}\,\,,
j_b=0\,\,,\n
{\bf div }{\cal F} &:=& [D_l, {\cal F}^l]_\star\,\,, 
({\bf rot {\cal F}})_l:=\epsilon_{lmn}[D_m, {\cal F}^n]_\star\,\,,\n
({\bf grad}\varphi)_l &:=& [D_l,\varphi]_\star\,\,, \dot{\cal F}:=[D_0, {\cal F}]_\star\,\,,\n
({\cal E}^l, {\cal B}^l) &:=& ({\cal F}_{0l}, 
1/2 \epsilon^{lmn}{\cal F}_{mn}) \cdots \mbox{written as } {\cal F}^l
\eeqa

%%%%%%%%%%%%%%%%%%%%%%%%%%%%%%%%%%%%%%%%%%%%%%%%
%%%%%%%%%%%%%%%%%%%%%%%%%%%%%%%%%%%%%%%%%%%%%%%%
%%%%%%%%%%%%%%%%%%%%%%%%%%%%%%%%%%%%%%%%%%%%%%%%
%%%%%%%%%%%%%%%%%%%%%%%%%%%%%%%%%%%%%%%%%%%%%%%%

\section{Electric-Magnetic Duality in  the Matrix Model}
\setcounter{equation}{0}

In this section
we consider the electric-magnetic duality of NCU(1) 
on a D3-brane in the matrix model.

First we summarize the duality.
When a theory can be written in two different ways
and just Legendre transformation  exchanges the two,
they are dual to each other. 
Namely in NCU(1)\cite{HL}\cite{RU}:
\beqa
Z=\int{\cal D}{\cal A}  e^{S_{NCU(1)}[{\cal A}, g_{YM}, \theta]}
=\int{\cal D}{\cal A}_D e^{S_{NCU(1)}[{\cal A}_D, g_{YM D}, \theta_D]}
\eeqa
where suffices $D$ means the dual.
The relation between them are:
\beqa
g_{YM D} &=& {1 \over g_{YM}}\,\,,\n
\theta^{\alpha \beta}_D
&=&{g_{YM}^2 \over 2}{\epsilon^{\alpha \beta}}_{\gamma \delta}\theta^{\rho \delta}\,\,,\n
{\cal F}_{\alpha \beta D}
&=&{1 \over 2 g_{YM}^2 }{\epsilon_{\alpha \beta}}^{\gamma \delta}{\cal F}_{\rho \delta} 
+ O(\theta)\,\,.\n
\label{dual0}
\eeqa
This is a spacetime-spacespace duality as well as electric-magnetic and strong-weak.

Next we will see this in our case.
By using eq.(\ref{gYM}), 
the dual Yang-Mills coupling and noncommutativity 
of eq.(\ref{dual0}) are written as:
\beqa
g^2_{YM D}&:=&{ 1 \over (2\pi g)^2 \theta^{01}\theta^{23}}\,\,,\n
\theta^{01}&:=&(2\pi g)^2 \theta^{01}(\theta^{23})^2\,\,,\n
\theta^{23}&:=&(2\pi g)^2 (\theta^{01})^2 \theta^{23}\,\,.
\label{dual}
\eeqa
Since there is a relation eq.(\ref{gYM}) in original theory, 
we would like the dual theory also to have the same one:
$g^2_{YM D}=(2\pi g_D)^2 \theta^{01}_D\theta^{23}_D$.
This requirement determine how the coupling $g$ changes to $g_D$:
\beq
g_D:={1 \over (2\pi)^4 g^3 (\theta^{01} \theta^{23})^2}
\label{dualg}
\eeq

We try to explain this by imaging a duality web.
The partition function of the matrix model is not changed 
under suitable rescaling of matrices.
That is, rescaling of coupling $g$ dose not change the model.
We have started with a $g$ 
and chosen an arbitrary back ground $\theta$.
On the other hand we can start by another $g_D$ and $\theta_D$,
and if those satisfy the condition:
\beq
(2\pi g)^2 \theta^{01}\theta^{23} \cdot (2\pi g_D)^2 \theta^{01}_D\theta^{23}_D=1\,\,,
\eeq
then two NCU(1)'s are the dual to each other.
This duality transformation is just rescaling: $g \rightarrow g_D$,
which is a symmetry.
It is natural to understand this 
if we remind type IIB superstring is self S-dual 
and its matrix model too.     
Since the matrix model is Morita equivalent(T-dual) to NCYM\cite{IIKK},
understanding of U-duality may clarify this duality web.

Finally let us see, in particular,  
more simple and familiar case.
We can find the dual pair of NCU(1)  from the matrix model with the same $g$,
and find the electric-magnetic duality for usual commutative Maxwell equations.
The U(1) coupling $g_{YM}$ cannot to be identity 
by rescaling of the gauge field ${\cal A}$ in noncommutative case. 
But, in the matrix model framework,  it is possible.
For given $g$, we choose the back ground solution with $\theta^{01}$ and $\theta^{23}$ which satisfy 
$(2\pi g)^2 \theta^{01}\theta^{23}=1$, namely, $g_{YM}=1$.
Then the dual noncommutativities also satisfy the same condition. 
Now the dual transformation is:
\beqa
(\theta^{01}_D,\theta^{23}_D) &=& (\theta^{23},\theta^{01})\,\,,\n
({\cal E}_D, {\cal B}_D ) &=& ({\cal B}, {\cal E} ) +O(\theta)\,\,.
\label{Maxdual}
\eeqa
Thus, Maxwell equations without the sources have a duality 
in  the following commutative limit: 
\beq
\theta, \theta_D \rightarrow 0\,\,,\,\,\, (2\pi g)^2 \theta^{01}\theta^{23}=1\,\,.
\eeq

%%%%%%%%%%%%%%%%%%%%%%%%%%%%%%%%%%%%%%%%%%%%%%%%%%%%%%%%%%%%%%%%%%%%%%%%
%%%%%%%%%%%%%%%%%%%%%%%%%%%%%%%%%%%%%%%%%%%%%%%%%%%%%%%%%%%%%%%%%%%%%%%%
%%%%%%%%%%%%%%%%%%%%%%%%%%%%%%%%%%%%%%%%%%%%%%%%%%%%%%%%%%%%%%%%%%%%%%%%
%%%%%%%%%%%%%%%%%%%%%%%%%%%%%%%%%%%%%%%%%%%%%%%%%%%%%%%%%%%%%%%%%%%%%%%%

\section{ Decoupling and Commutative Limit}
\setcounter{equation}{0}

In this section 
we are going to define a decoupling limit( dimensional reduction limit) and
a commutative limit in the matrix model to push the brane world scenario. 
We assume the spacetime dimension is almost equal to 4 
and coordinates are almost commutative. 
It dose not have to be exact 4 dimensional commutative spacetime.
The stand point of this paper is in 10 dimensional noncommutative one.  
Our strategy is getting the above brane world from the matrix model in 10 dimension,
by fine tuning the parameters $\theta^{\mu \nu}\,\,,(\mu, \nu =0 \sim 9)$.

We define the decoupling limit as:
\beq
\theta^{\mu \nu} \rightarrow \infty\,\,.
\label{decouple}
\eeq
In this limit, 
\beq
[\hat{p}_\mu, \hat{p}_\nu ] \rightarrow 0\,\,.
\eeq
This means the dimension of brane get down by two in $\mu \nu$ directions, 
so it is natural definition.

To explain this by string terminology, 
we need identify strings in the matrix model.  
It is possible\cite{IKK}\cite{IIKK}.
We will summarize it in the non selfdual case.
The von Neumann lattice is the best representing the intuitive spacetime \cite{IKK}\cite{IIKK}\cite{GN}.   
It is constructed by using coherent state of operators of the noncommutative coordinates
which forms Heisenberg algebra: eq(\ref{xx}). 
The lattice spacing is $\sqrt{2\pi \theta^{01}}$ for $0,1$ directions 
and  $\sqrt{2\pi \theta^{23}}$ for $2,3$, 
which are written as $l^{\mu}_{NC}$.
Because of the noncommutativity,  
states cannot be localized.
So, fields are naturally represented as bi-local ones,
which are functions of two points.
Small momentum modes, namely, 
the first and third line of eq.(\ref{momenta})
correspond ordinary (commutative) field.
Large momentum modes, the second and forth line of eq.(\ref{momenta}) 
correspond open strings. 
Define $d^\mu:=\theta^{\mu \nu} k_\nu$ 
and decompose $d$ as $d=d_0 + \delta d$, 
where $d_0$ is a vector 
which connects two points on the lattice and $|\delta d^\mu| < l^{\mu}_{NC}$.
The length of open string is $d_0^\mu$
and the momentum which can be associated with the center of mass motion of open string 
is ${k_c}_\mu:=(1/\theta)_{\mu \nu} \delta^{\nu}d$. 
There is an inequality:
\beq
 |{k_c}_\mu| < \sqrt{2\pi \over |\theta^{\mu-1, \mu}|}\,\,.
\eeq 
In the decoupling limit eq.(\ref{decouple}), 
only open strings survive 
(see eq.(\ref{momenta}) where n assumed to be large enough).
The momenta of the open strings $k_c$ goes to zero.
If one considers the higher order correction to propergator of bi-local field,
then the oscillation of open string are seen 
by collecting open strings bits( see fig.).
In the limit, however, the momentum of the bits are goes to zero 
and the open string cannot make loop 
and is decoupled from closed strings in bulk.
So we call this limits as noncommutative open string( NCOS) limit.
In the paper\cite{IIKK} they identified 
the effective tension of the open string in the matrix model 
to the noncommutativity:
\beq
T_{eff}=1/\theta\,\,,
\eeq 
Our decoupling limits means the tension of the open strings goes to zero.

These features are completely parallel 
to string theory approach by the world sheet with NS-NS B filed\cite{SST}\cite{GMMS}.
In order to see this clearly
we can use the relation 
between string theory and the matrix model. 
In the paper\cite{KT} we have gotten the relation 
between  IIB open string with a D-brane 
having both electric and magnetic field on the D-brane   
and the matrix model solution having both spacetime and spacespace noncommutativity.
In fact, we compared the interactions of two Dp-branes 
with various charges of lower dimensional branes in different two approaches.
Since the matrix model is defined in Euclid signature 
the result should be wick rotated after the calculation\cite{M}.
Electromagnetic field $\tilde{E}$, $\tilde{B}$ on the D-brane in string theory 
and noncommutativity in matrix model are related as$^{}$\footnote
{In the paper\cite{KT}, we used flat metric. To see the dependence of diagonal metric
diag$(-g_e, g_e, g_b, g_b)$ we regard $\tilde{E}=\tilde{E}_{\small flat}/g_e $ etc.  
}:
\beqa
(1-\tilde{E}^2){\theta^{01}}^2&=&\left( 2 \pi \alpha^\prime \right)^2\,\,,\n
(1+\tilde{B}^2){\theta^{23}}^2&=&\left( 2 \pi \alpha^\prime \right)^2\,\,.
\label{KT2}
\eeqa
Noncommutativities in the matrix model and in string theory are different from each other,
and above equations(\ref{KT2}) tells us the relation to $\theta^{01,23}_{s}$ in string theory$^{}$\footnote
{We use notations in Ref.\cite{SST}, where $\theta=\theta^{01}_s, g=g_e$ here.}:
\beqa
(\theta^{01})^2 &=& {(2\pi \alpha^\prime)^2 \over 2}\left[
\sqrt{1+\left(2 g_e \theta^{01}_s \over  2\pi \alpha^\prime \right)^2}
+1
\right]\,\,,\n
(\theta^{23})^2 &=& {(2\pi \alpha^\prime)^2 \over 2}\left[
\sqrt{1+\left(2 g_b \theta^{23}_s \over  2\pi \alpha^\prime \right)^2}
-1
\right]\,\,,
\label{thetas}
\eeqa
where the closed string metric is written as diag$(-g_e, g_e, g_b, g_b)$.
The critical electric field limit(NCOS limit in string theory) is 
\beqa
|\tilde{E}| &\rightarrow& 1\,\,,\n
\alpha^\prime &:& \mbox{fixed}\,\,,\n
g_e &\sim& {1 \over 1-\tilde{E}^2}\,\,.
\eeqa
Then noncommutativities tend to:
\beq
\theta^{01} \rightarrow \infty\,\,,\,\,\, \mbox{while}\,\, \theta^{01}_s: \mbox{finite}\,\,,
\eeq
and consistent with string theory approach.
%$^{}$\footnote
%{On the other hand in the NC field theory limits correspond to: $ \theta^{01} \rightarrow ?$.}.

Now, to  see the connection to decoupled case,
we may represent the solution $\hat{p}_\alpha$ in eq.(\ref{sol}) 
as commutative $ {\hat{p}^{comm}}_\alpha $'s and ${\hat{q}_{comm}}^\alpha$'s:
\beqa
&&\hat{p}_\alpha={\hat{p}^{comm}}_\alpha + {1 \over 2}F_{\alpha \beta }{\hat{q}_{comm}}^\beta\,\,,\n
&&\left[{\hat{p}^{comm}}_\alpha, {\hat{q}_{comm}}^\beta \right]=i \delta_\alpha^\beta\,\,.
\eeqa
Other commutators are equal to zero$^{}$\footnote
{This relation is, in fact, satisfied when we represent $\hat{p}^{comm}$, $\hat{q}_{comm}$ as:
\beqa
{\hat{p}^{comm}}_\alpha&=&1 \otimes 1 \otimes \cdots \otimes \hat{p} \otimes \dots \,\,,\n
{\hat{q}_{comm}}^\alpha&=&1 \otimes 1 \otimes \cdots \otimes \hat{q} \otimes \dots \,\,,
\eeqa
where $\hat{p}$ and $\hat{q}$ are in $\alpha$'s place, 
and are satisfy $[\hat{p},\hat{q}] = i$.}.
So these ${\hat{p}^{comm}}$ are regarded 
as commutative background solution of the matrix model.
This representation clarify the relation 
to the lower dimensional brane solution in the decoupling limit.
Since $F=1/\theta$ the decoupling limit $\theta^{\alpha \beta} \rightarrow \infty$ 
means $F_{\alpha \beta} \rightarrow 0$.
When $F_{\alpha \beta}$ is zero, $(\alpha, \beta)$ direction cannot form the brane.  

Let us think about the opposite limit:
\beq
\theta^{\mu \nu} \rightarrow 0\,\,.
\label{commute}
\eeq  
Since 
\beq
[x^\mu ,x^\nu]_\star =-i\theta^{\mu \nu}\,\,,
\eeq
coordinates commute to each other in this limit.
The momentum $k_\mu$ is eigenvalue of $\hat{P}_\mu 
:=[\hat{p}, \mbox{\hspace{1mm}$\cdot$}\hspace{2mm}]$ (not of $\hat{p}$).
Momenta commute to themselves without any limits 
because $[\hat{P}_\mu, \hat{P}_\nu]=0$.
So the limit eq.(\ref{commute}) can be called the  commutative limit.
 
Next we try to draw a scenario of getting an almost commutative near 4 dimensional spacetime.
We start from, for instance, 6 dimensional solution of the matrix model.
There are three noncommutativity parameters $\theta^{01}, \theta^{23}, \theta^{45}$.    
We  think of regions near following limits:
\beqa
\theta^{01} &\rightarrow& 0 \,\,,\n 
\theta^{23} &\rightarrow& 0\,\,,\n
\theta^{45} &\rightarrow& \infty\,\,.
\eeqa
Then we have a 4 dimensional commutative spacetime.
Of cause this is not dynamical determination, but just a fine tuning.
That is beyond the scope of this paper. 
We will discuss this possibility, however, in the final section.

\begin{center}
\setlength{\unitlength}{3947sp}%
\begingroup\makeatletter\ifx\SetFigFont\undefined%
\gdef\SetFigFont#1#2#3#4#5{%
  \reset@font\fontsize{#1}{#2pt}%
  \fontfamily{#3}\fontseries{#4}\fontshape{#5}%
  \selectfont}%
\fi\endgroup%
\begin{picture}(2949,1260)(1489,-2311)
\thinlines
\put(1501,-1711){\line( 5, 2){750}}
\put(2251,-1411){\line( 5,-2){750}}
\put(3001,-1711){\line( 3, 1){832.500}}
\put(3826,-1411){\line( 1,-1){600}}
\put(2176,-1186){\makebox(0,0)[lb]{\smash{\SetFigFont{12}{14.4}{\rmdefault}{\mddefault}{\updefault}$k_c$}}}
\put(3001,-2086){\makebox(0,0)[lb]{\smash{\SetFigFont{12}{14.4}{\rmdefault}{\mddefault}{\updefault}$k_c^\prime $}}}
\put(3751,-1186){\makebox(0,0)[lb]{\smash{\SetFigFont{12}{14.4}{\rmdefault}{\mddefault}{\updefault}$k_c^{\prime\prime}$}}}
\put(4351,-2311){\makebox(0,0)[lb]{\smash{\SetFigFont{12}{14.4}{\rmdefault}{\mddefault}{\updefault}$k_c^{\prime\prime\prime}$}}}
\put(100, -2700){\makebox(0,0)[lb]{\smash{\SetFigFont{12}{14.4}{\rmdefault}{\mddefault}{\updefault}
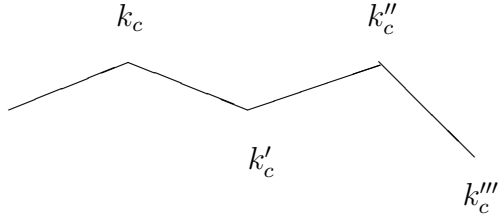
$\mbox{ a open string consisting of four string bits which have momenta $k_c, \cdots,   k_c^{\prime\prime\prime}$.}$}}}
\end{picture}
\end{center}
\vspace*{1cm}
%%%%%%%%%%%%%%%%%%%%%%%%%%%%%%%%%%%%%%%%%%%%%%%%%%%%%%%%%%%%%%%%%%%%%%%%
%%%%%%%%%%%%%%%%%%%%%%%%%%%%%%%%%%%%%%%%%%%%%%%%%%%%%%%%%%%%%%%%%%%%%%%%
%%%%%%%%%%%%%%%%%%%%%%%%%%%%%%%%%%%%%%%%%%%%%%%%%%%%%%%%%%%%%%%%%%%%%%%%
%%%%%%%%%%%%%%%%%%%%%%%%%%%%%%%%%%%%%%%%%%%%%%%%%%%%%%%%%%%%%%%%%%%%%%%%

\section{Conclusions and Discussions}
\setcounter{equation}{0}

We have considered the electric-magnetic duality 
and the decoupling-commutative limit in the matrix model.
4 dimensional spacetime can be constructed as D3-brane solution of the model.
It has non-selfdual solutions which we have treated in this paper. 
Then there are two non commutativity parameters: $\theta^{01}$, $\theta^{23}$. 

Electric-magnetic duality transformation changes those parameters 
as eq.(\ref{dual}),
as well as Yang-Mills coupling and electromagnetic fields. 
In addition, it corresponds to the rescaling of matrices 
in the original matrix model,
which has S-duality symmetry.
In particular solution related to a $g$, 
its duality is just the duality of U(1) Maxwell theory: eq.(\ref{Maxdual}).   

Decoupling limits have been defined as eq.(\ref{decouple}).
Open strings are decoupled from closed strings by looking into their momenta,
and the tension goes to zero.
This has been also seen from the relation $\theta^{01}$ and $\theta^{01}_s$ 
clearly in eq.(\ref{thetas}) 

Noncommutativity parameters manage 
making our 4 dimensional commutative spacetime.
When $\theta^{ij} \rightarrow \infty$ corresponding $i,j$ directions decouple 
while $\alpha \beta $ direction commutes as  $\theta^{\alpha \beta} \rightarrow 0$.
Staring from higher dimensional brane, 
we can get an  almost commutative and near 4 dimension spacetime,
by fine tuning of those parameters.
Changing parameters may be understood 
from the NCYM around the classical brane solution.
That is, bi-local fields on the brane may condense and change the back ground spacetime.
We write:
\beq
{\cal A}_\alpha(x) ={\cal A}_\alpha^0(x) + {\cal A}_\alpha^1(x)\,\,,
\eeq  
where ${\cal A}^0(x)$ is a solution of NCU(1).
In the matrix model, we can regard ${\cal A}^0(x)$ as a part of back ground.
Namely, new background happens: 
\beq
\hat{p}_\alpha  \rightarrow \hat{p}_\alpha^\prime = \hat{p}_\alpha + \hat{\cal A}^0_\alpha\,\,. 
\eeq
From this expression we find the spacetime and the field on it are treated unified way. 
For example, 
if we choose a solution of eq.(\ref{Maxwell}):
${\cal A}^0_\alpha(x)={1 \over 2}{\cal F}^\prime_{\alpha \beta} x^\beta$,
(${\cal F}^\prime$: constant),
then $\hat{p}^\prime$ satisfies:
\beq
\left[\hat{p}^\prime_\alpha, \hat{p}^\prime_\beta \right]=i (F+{\cal F})_{\alpha\beta}\,\,,
\eeq
where ${\cal F}$ is field strength of ${\cal A}^0$ and relates to ${\cal F}^\prime$ 
as ${\cal F} = {\cal F}^\prime + {1 \over 4} {\cal F}^\prime  F^{-1} {\cal F}^\prime$.
Thus the background spacetime can change dynamically. 
So, considering solutions of NCYM 
may give understanding of the relation to the commutative (different dimensional) background\cite{IK}, 
S-dual back ground, and so on. 

Recently there are also various studies for 
nonperturbative solution of NCYM:  
\cite{NS}\cite{Furuuchi}\cite{GN} \cite{NV}\cite{GN2}.  
It is interesting to map them to the matrix in the sense of: ${\cal A}^0 \rightarrow \hat{\cal A}^0$. 
In this paper 
we have concentrated to the electric-magnetic duality,
only because of avoiding complexity in the first step.
We have seen, however, full S-duality with supersymmetry 
may have more interesting feature,
combining T duality. 
NCYM from IIB string theory are studying now by a lot of group 
and F-string and D-string are considered there\cite{SST}\cite{GMMS}\cite{LRS}\cite{RU}.
It is also possible to consider them 
in the matrix model, which is going to be our next theme. 
With the help of above related approaches,
we probably understand, in near future, how our 4 dimensional universe happens.

\begin{center} \begin{large}

Acknowledgments
\end{large} \end{center}
I am grateful for stimulative situations in IMSc.

\setcounter{equation}{0}

\end{document}